\documentclass[traditabstract]{aa}  
\usepackage{graphicx}
\usepackage{txfonts}
\usepackage{natbib}
\bibpunct{(}{)}{;}{a}{}{,}
\begin{document}
   \title{Probing the evolution of molecular cloud structure}

   \subtitle{II: From chaos to confinement}

   \author{J. Kainulainen
          \inst{1},
           H. Beuther
          \inst{1},
           R. Banerjee
          \inst{2},
           C. Federrath 
          \inst{2, 3},
%          \inst{2}
%          \inst{2}
 %         \inst{1}
 \and
           T. Henning
          \inst{1}
%           M.-M. Mac Low
%          \inst{4}
%           R. Plume
%          \inst{3}
          }

   \offprints{jtkainul@mpia.de}

   \institute{Max-Planck-Institute for Astronomy, K\"onigstuhl 17, 69117
     Heidelberg, Germany \\
              \email{[jtkainul;beuther;henning]@mpia.de}
         \and
  Zentrum f\"ur Astronomie der Universitat Heidelberg, Institut f\"ur Theoretische Astrophysik, 69120 Heidelberg, Germany
         \and
Ecole Normale Sup\'{e}rieure de Lyon, CRAL, 69364 Lyon, France
%             AMNH \\
%         \and
%             Univ. Calgary \\
            }

   \date{Received ; accepted }
% \abstract{}{}{}{}{} 
% 5 {} token are mandatory 
 %  {We have recently studied the distribution of column densities in molecular clouds. }
  % {In this paper, we examine the column density structure of molecular clouds using near-infrared dust extinction data of nearby ($r \lesssim 500$ pc) molecular clouds. In particular, we consider the stability of cloud structures and the connection of  processes responsible for shaping clouds at low column densities.}
%   {We demonstrate that molecular clouds show a structural transition at a regime characterized by the surface mass density of $\Sigma \approx 40-80$ M$_\odot$ pc$^{-1}$ and length scales of @@@. The transition can be understood as a jump from diffuse, transient structures to structures confined by the external pressure invoked by the diffuse medium surrounding them.  }
   %Traditional abstract:
\abstract{We present an analysis of the large-scale molecular cloud structure and of the stability of clumpy structures in nearby molecular clouds. % We present a new approach to analysing large-scale molecular cloud structure, and examine the stability of clumpy structures in molecular clouds.
In our recent work, we identified a structural transition in molecular clouds by studying the probability distributions of gas column densities in them. In this paper, we further examine the nature of this transition. The transition takes place at the visual extinction of $A_\mathrm{V}^\mathrm{tail} = 2-4$ mag, or equivalently, at $\Sigma^\mathrm{tail} \approx 40-80$ M$_\odot$ pc$^{-2}$. The clumps identified above this limit have wide ranges of masses and sizes, but a remarkably constant mean volume density of $\overline{n} \approx 10^3$ cm$^{-3}$. This is $5-10$ times larger than the density of the medium surrounding the clumps. By examining the stability of the clumps, we show that they are gravitationally unbound entities, and that the external pressure from the parental molecular cloud is a significant source of confining pressure for them. Then, the structural transition at $A_\mathrm{V}^\mathrm{tail}$ may be linked to a transition between this population and the surrounding medium. The star formation rates in the clouds correlate strongly with the total mass in the clumps, i.e, with the mass above $A_\mathrm{V}^\mathrm{tail}$, dropping abruptly below that threshold. These results imply that the formation of pressure confined clumps introduces a prerequisite for star formation. Furthermore, they give a physically motivated explanation for the recently reported relation between the star formation rates and the amount of dense material in molecular clouds. Likewise, they give rise to a natural threshold for star formation at $A_\mathrm{V}^\mathrm{tail}$.}

  \authorrunning{J. Kainulainen et al.}
  \titlerunning{From chaos to confinement}
  \maketitle

%________________________________________________________________

%*************************
%*************************
\section{Introduction} %*
%*************************
%*************************
\label{sec:intro}

% Starting paragraph

Formation of dense, self-gravitating structures inside more diffuse, large-scale molecular clouds is the ultimate prerequisite for star formation. In addition to self-gravitating dense cores, molecular clouds in which star formation is taking place show exhaustive structural complexity characterized by large contrasts in both density and velocity. From the general observation that almost all known molecular clouds harbor young stars, it is known that the formation of structures capable of star formation (or alternatively, cloud dissipation) must proceed relatively rapidly compared to the life-times of molecular clouds. Likely as a result of this complexity and rapid development, molecular clouds also show wide ranges of star-forming efficiencies and -rates \citep[e.g.][]{hei10, lad10}. This connection between the cloud structure and the capability of a cloud to form stars makes determining the roles of processes and parameters that control the cloud structure a fundamental open topic in the physics of star formation \citep[reviewed, e.g., by][]{mck07, mac04}. 

% This is further signified by the fact that the few known non-star-forming clouds do not seem to harbor cores that reach densities typically observed in the dense cores harbored in star-forming clouds \citep[e.g.][]{lom06}. These non-star-forming clouds are, however, not structureless either but often show the imprints of structures like clumps and filaments, i.e. seemingly simply lacking the dense cores in them \citep[e.g.][]{lom06, war10}. Thus, it clearly appears that the 

% Structures in clouds.

%The structure formation in molecular clouds is strongly affected, and possibly also driven, by supersonic turbulent motions at the scales larger than molecular clouds. The energy transfer from these scales downwards induces a scale-dependent cascade of kinetic energies, where the velocity differences between  (a proxy of kinetic energy)

% To the point: the turbulent structure of clouds

   In the current analytic models of star formation, one particularly important structural parameter of molecular clouds is the probability density function (PDF, hereafter) of volume densities, which describes the probability of a volume $dV$ to have a density
   between $[\rho,\rho + d\rho]$. In such theories, the function has pivotal role: it is used to explain among others
   the initial mass function of stars, and the star formation
   rates and efficiencies of molecular clouds \citep[e.g.][]{pad02, kru05, elm08, hen09}. In particular, this distribution is expected to take a
   log-normal shape in isothermal, turbulent media not significantly affected by the self-gravity of gas
   \citep[e.g.][]{vaz94, pad97, ost99, fed08b}. Most importantly from the observational point-of-view, the log-normality of the distribution is expected to be reflected in the probability distributions of \emph{column densities} in molecular clouds \citep{vaz01, goo09, fed10}. Also, recently a method has been developed to attain information of the actual three-dimensional density PDF based on the observed, two-dimensional column density PDFs \citep{bru10new, bru10, bru10taurus}. Even though it has been pointed out that the general log-normal-like form for the (column) density PDF can be borne out by various processes \citep{tas10}, it is obvious that an accountable theory of cloud structure must meet with the observed characteristics of the distribution. This is particularly the case if the probability distribution shows any scale-dependent features and/or time evolution. Such properties have indeed been predicted, e.g. in the presence of strong self-gravity \citep{kle00, fed08a, cho10, kri10}, and scale-dependent features have also been recently observed \citep{kai09, fro10, pin10jorge}. This makes probing column density probability distributions one measure of cloud structure that can be used to constrain analytic star formation theories.

% Observational work 

%[@@@  Elmegreen \& Scalo (2004)]
%[@@@ Krumholz, Matzner \& McKee (2006)]

However, the connection between theoretical and numerical predictions with observations of the PDF has been poorly investigated. The studies in which the column density PDFs of mostly individual clouds have been examined have found a qualitative agreement with the predicted log-normal shape \citep[e.g.,][]{rid06b, goo09, but09}. The lack of systematic studies of column density PDFs has been mostly due to observational obstacles: all observational tracers of the cloud mass distribution suffer from shortcomings specific to the tracer in question \citep[see, e.g.,][]{goo09}. Generally, the dynamical ranges probed by different molecular emission line tracers are often narrow, thereby probing only a limited range of the PDF. Dust continuum emission observations probe a wider dynamical range of column densities, but become insensitive at column densities below $N \lesssim \mathrm{a\  few\ } \times 10^{21}$ cm$^{-2}$, thus missing a regime where most of the cloud mass is. This is the case also for dust extinction measurements using infrared shadowing features. In addition to these restrictions, mapping nearby cloud complexes that often span several degrees on the sky at high sensitivity requires a colossal observational effort, not generally feasible through typical observing campaigns. Dust extinction mapping in the near-infrared reaches only modest column densities of $N \lesssim \mathrm{a\  few\ } \times 10^{22}$ cm$^{-2}$, thereby mostly missing dense star-forming clumps and cores. However, near-infrared extinction mapping reaches very efficiently the low column densities $N \sim \mathrm{1-3} \times 10^{22}$ cm$^{-2}$ \citep{lom01}, a regime where most of the cloud mass is. Therefore, it provides a feasible tool to measure the column density PDFs at the scales of entire cloud complexes. The method has indeed been used recently for this purpose, especially by \citet{kai09} \citep[see also][]{lom06, lom08b, fro10, lom10, pin10jorge}).

% Our work

In our recent work \citep[][Paper I hereafter]{kai09}, we presented the first systematic study of the column density PDFs in all nearby molecular clouds closer than 200 pc. We used near-infrared dust extinction maps of $23$ molecular clouds to identify a transition in the PDF shape from a log-normal-like shape at lower column densities to a power-law-like shape at higher column densities. Such transition is characteristic to all star-forming molecular clouds. However, we showed that some of the non-star-forming clouds in our sample did not have the transition, but their PDFs were well fitted by a log-normal over the entire range of column densities above the detection limit. This led us to speculate that the PDF feature is linked to a transition from a quiescent regime dominated by turbulent motions to a regime of active star formation dominated by gravity. The non-star-forming clouds not showing the feature would then belong entirely to the former regime, and during their subsequent evolution towards star formation gravitationally dominated structures would appear, inducing also a transition to the shape of their PDFs.

In this paper, we present a more detailed analysis of the structural transition identified from the column density PDFs. In particular, we will examine the physical characteristics and stability of the structures identified using the PDFs. With this analysis, we will show that the change in the PDF shape can be understood as a transition between the diffuse, interclump medium and a population of clumps that are gravitationally unbound, but significantly supported against dispersal by the external pressure imposed to them by the surrounding medium. This interpretation links the PDF shape to a physically motivated explanation for the relation between star formation rates and the amount of high-density material in molecular clouds reported recently by \citet{lad10}, and also to the threshold of star formation in molecular clouds.

% In this paper.

In \S\ref{sec:data} we shortly describe the dust column density data used in this paper. In \S\ref{sec:results} we characterize the structures identified from the column density maps and examine their stability. In \S\ref{sec:discussion} we discuss the results and their impact for the structure- and star formation in molecular clouds . In \S\ref{sec:conclusions} we give our conclusions. 

% [@@@ Di Francesco 2007 (cores), Ward-Thompson et al 2007 for clump+core properties!]

%****************************************************************
%****************************************************************
\section{The column density data of nearby clouds}                     %*
%****************************************************************
%****************************************************************
\label{sec:data}

In Paper I, we used the near-infrared color-excess mapping method presented by \citet{lom09} \citep[see also][]{lom05, lom01}, namely \textsf{nicest}, to derive dust extinction maps for 23 nearby molecular clouds. The technique was used in
conjunction with near-infrared data from the 2MASS survey \citep{skr06}, resulting in dust column density maps covering the dynamical range of $A_\mathrm{V} \approx 1.2-25$ mag in the spatial resolution of 0.1 pc ($\sim 2\arcmin$ at the distance of the nearby clouds). While the cloud sample for this paper is otherwise the same as in Paper I, we have excluded the Coalsack cloud from the analysis. Our recent molecular line observations of the Coalsack have shown that the region likely includes a significant extinction component not only from Coalsack, but also from an extended cloud at a larger distance (Beuther et al., in prep.). Since the effect of that component may well disturb the statistics derived in this paper, we decided to exclude Coalsack from the sample.

\begin{figure*}
   \centering
   \includegraphics[bb=50 35 500 315, clip=true, width=0.99\textwidth]{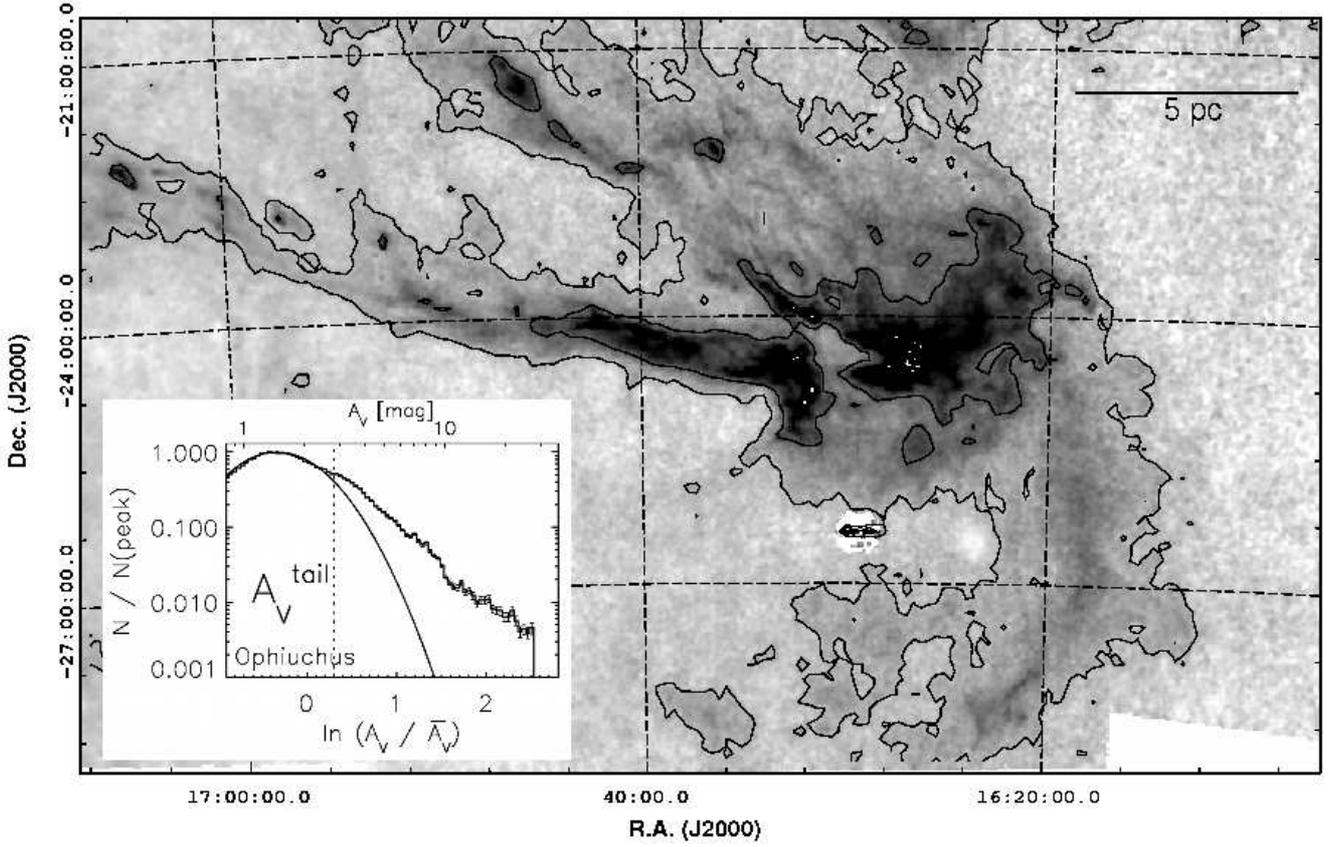} 
      \caption{Dust extinction through the Ophiuchus molecular cloud, derived using the near-infrared extinction mapping technique and 2MASS data (Paper I). The contours are drawn at $A_\mathrm{V}^\mathrm{tail} = 2.8$ mag, and at $A_\mathrm{V} = 1$ mag which is used to calculate the total mass of the cloud. The mass in the dense component is $\sim 21$ \% of the mass above $A_\mathrm{V} = 1$ mag in this cloud (cf. Table \ref{tab:clouds}). The inset shows the column density PDF of the cloud, with an approximate log-normal function fitted to the peak of it.
              }
         \label{fig:map-ophiuchus-pdf}
   \end{figure*}

As an example of our data, Figure \ref{fig:map-ophiuchus-pdf} shows the extinction map derived for the Ophiuchus cloud. The figure shows also the PDF of the cloud, with the lognormal-like low-$A_\mathrm{V}$ part and the power-law-like tail at high-$A_\mathrm{V}$ clearly separable. The transition between these parts occurs approximately at $A_\mathrm{V}^\mathrm{tail} \approx 2.8$ mag in this cloud. Throughout this paper, we refer to such position in the PDFs with $A_\mathrm{V}^\mathrm{tail}$. In the clouds included in the study, the transition occurs at relatively low $A_\mathrm{V}$ values, $A_\mathrm{V}^\mathrm{tail} = [2.0, 11]$ mag, although in most cases between $A_\mathrm{V}^\mathrm{tail} = 2-4$ mag. The $A_\mathrm{V}^\mathrm{tail}$ values determined for each cloud are listed in Table \ref{tab:clouds}. 

The $A_\mathrm{V}^\mathrm{tail}$ value defines a set of spatially closed iso-contours in the column density maps (see Fig. \ref{fig:map-ophiuchus-pdf}). Throughout the paper, we will refer to the region where $A_\mathrm{V} \lesssim A_\mathrm{V}^\mathrm{tail}$ as the \emph{diffuse component}, and similarly, to all regions where $A_\mathrm{V} \gtrsim A_\mathrm{V}^\mathrm{tail}$ as the \emph{dense component}. The former refers then, by definition, to the log-normal part of the PDF and the latter to the power-law-like part. Morphologically, the diffuse component is a uni-body structure in all complexes, but the dense component forms separate structures. We will refer to such separate structures as \emph{clumps} in this paper.

%**************************
%**************************
\section{Results}           %*
%**************************
%**************************
\label{sec:results}

In this section, we use the column density data introduced in \S\ref{sec:data}
to examine the nature of the diffuse and dense components. We first
examine the physical characteristics of the components, namely the masses and sizes, densities, and velocity structure. Then, we consider the observed characteristics from the point of view of pressure balance in molecular clouds.

%************************************************************************
\subsection{Characteristics of the diffuse and dense components}
%************************************************************************
%
%------------------------------------
\subsubsection{Total mass}
%------------------------------------
\label{sec:mass}

% Mass ratio of the components

The gas column densities in the clouds can be inferred from the extinction maps using the measured extinction-to-gas column ratio. We transformed the visual extinction values in the maps to hydrogen column densities using the relation \citep{boh78}:
\begin{equation}
\beta = (N(H) + 2N(H_2))/A_\mathrm{V} = 9.4 \times 10^{20} \mathrm{\  cm}^{-2} \mathrm{\  mag}^{-1}.
\label{eq:bohlin}
\end{equation}
We then calculated the total mass of the cloud as a sum of extinction values above $A_\mathrm{V} = 1$ mag:
\begin{equation}
M_\mathrm{tot} = D^2 \mu \beta \times \int_{\Omega : A_\mathrm{V} > 1 \mathrm{\  mag}}{A_\mathrm{V}\   \mathrm{d}x \mathrm{d}y},
\label{eq:mass_tot}
\end{equation}
where $D$ is the distance to the cloud, $\mu=1.37$ is the mean molecular weight (adopting the same values as \citet{lom08b}, i.e., 63 \% hydrogen, 36 \% helium, and 1 \% dust), and $x$ and $y$ refer to the map pixels. We adopt the same distances for clouds as listed in Paper I. We note that the chosen lower limit of $A_\mathrm{V} = 1$ mag is arbitrary, and the total mass depends on the selected value (choosing lower threshold will yield higher masses for all clouds). However, a fixed value will make the values comparable between the clouds. We also note that the $A_\mathrm{V} = 1$ mag contour is closed in most mapped regions, thus uniformly defining a cloud boundary. Table \ref{tab:clouds} lists the mean extinctions, $\overline{A}_\mathrm{V}$, for the clouds calculated using this definition for a cloud.

The total mass of the dense component was calculated from the extinction in excess to the $A_\mathrm{V}^\mathrm{tail}$ threshold level:
\begin{equation}
M_\mathrm{dense} = D^2 \mu \beta \times \big[\int_{\Omega : A_\mathrm{V} > A_\mathrm{V}^\mathrm{tail}}{A_\mathrm{V}\   \mathrm{d}x \mathrm{d}y} - A_\mathrm{V}^\mathrm{tail} \times \int_{\Omega : A_\mathrm{V} > A_\mathrm{V}^\mathrm{tail}}{\mathrm{d}x \mathrm{d}y}\big],
\label{eq:mass_dense}
\end{equation}
The mass of the diffuse component was then defined as $M_\mathrm{diffuse} = M_\mathrm{tot} - M_\mathrm{dense}$. We list in Table \ref{tab:clouds} the ratios of the mass of dense component to the total mass of the cloud. Clearly in all clouds, the mass of the diffuse component dominates the cloud mass. The ratios vary from a few percents for clouds with low star-forming activity to $\sim$20 \% for the most active clouds. Note that the total mass of the cloud was calculated as the mass above $A_\mathrm{V} > 1$ mag. Since the column density below this level is, of course, not zero, our total masses represent lower limits. Accordingly, the quoted $M_\mathrm{dense} / M_\mathrm{tot}$ ratios represent upper limits.

\begin{table}
\begin{minipage}[t]{\columnwidth}
\caption{Molecular clouds and the derived properties.}
\centering
\renewcommand{\footnoterule}{}  % to avoid a line before footnotes
\begin{tabular}{lcccccc}
\hline \hline
Cloud & $A_\mathrm{V}^\mathrm{tail}$ & $A_\mathrm{V}^\mathrm{90}$\footnote{Extinction above which the contribution of the log-normal component to the PDF is less than 10 \%.} & $\overline{A}_\mathrm{V}$\footnote{Calculated using the column density values above $A_\mathrm{V} > 1$ mag.} & $M_\mathrm{tot}$ [10$^4$ M$_\odot$]$^b$ & $\frac{M_\mathrm{tail}}{M_\mathrm{tot}}$ & $N_\mathrm{clumps}$ \\ 
\hline
\multicolumn{6}{l}{Star-forming clouds, physical resolution 0.1 pc}\\
\hline                      
\object{Ophiuchus}    	&  2.8 	& 6.7		& 2.4 & 0.52 & 0.21 & 20 \\ 
\object{Taurus}           	&  4 		& 7.7		& 2.2 & 0.99 & 0.09 & 33 \\
\object{Serpens}\footnote{$A_\mathrm{V}^\mathrm{tail}$ could not be defined properly for the cloud.}            	&  - 	& -		& - 	 & - 	    & -  		&  -\\
\object{Cha I}            		&  2.1 	& 3.2		& 1.7 & 0.27 & 0.16 & 11\\
\object{Cha II}           		&  2.1 	& 3.2		& 1.8 & 0.11 & 0.17 & 7  \\
\object{Lupus III}        	&  3 		& 4.1		& 1.7 & 0.08 & 0.07 &  18 \\
\object{CrA cloud}        	&  2.3 	& 4.1		& 1.7 & 0.10 & 0.13  & 11 \\
\object{Lupus I}          	&  2.0 	& 3.3		& 1.6 & 0.13 & 0.12 &  16\\
\object{LDN1228}\footnote{The most prominent Lynds Dark Nebula in the region.}     &  2.1 	& 3.1		& 1.5 & 0.15 & 0.09 & 4 \\	
\object{Pipe}$^c$          	&  - & -		& - & - 		& -		&  -\\ 
\object{LDN134}$^d$        &  2.2 	& 3.7		& 1.5 & 0.09 & 0.07 & 3 \\
\object{LDN204}$^d$
					&  3.2 	& 5.3		& 1.7 & 0.26 & 0.04 &  10\\
\object{LDN1333}$^d$	&  3.1 	& 5.0		& 1.5 & 0.25 & 0.01 &  9 \\
\hline
\multicolumn{5}{l}{Non-star-forming clouds, physical resolution 0.1 pc}\\
\hline                      
\object{LDN1719}$^b$	&  - 		& -	& 1.8   & 0.18 &  - & - \\	
\object{Musca}            	&  2.0 	& 3.2 	& 1.7 & 0.04 & 0.16 & 3 \\
\object{Cha III}          		&  2.1 	& 5.5		& 1.7 & 0.12 & 0.11 & 26 \\
\object{Lupus V}          	&  - 		& -	& 1.8 & 0.28 &  - & - 	\\
\hline 
\multicolumn{5}{l}{Star-forming clouds, physical resolution 0.6 pc}\\
\hline                      
\object{Ori A GMC}        	&  3 		& 5.6		& 2.2 & 9.2 & 0.20 &  15\\  	
\object{Per cloud}        	&  3 		& 4.9		& 2.0 & 1.4 & 0.12  & 10 \\	
\object{Ori B GMC}        	&  2.1 	& 5.4		& 1.9 & 7.0 & 0.19 &  35\\	
\object{Cepheus A}        	&  3.6 	& 13		& 2.7 & 1.1 & 0.10 & 3  \\ 
\object{California}       	&  4.2 	& -		& 1.7 & 10  & 0.01 & 14  \\ 	
\end{tabular}
\label{tab:clouds}
\end{minipage}
\end{table}

%--------------------------------------------------------------------
\subsubsection{Clumps in the dense component}
%--------------------------------------------------------------------
\label{sec:size}

In the following, we characterize the individual structures, i.e. clumps, in the dense component. We used a simple thresholding approach to identify the clumps from the extinction maps, namely the "clumpfind2d" routine \citep{wil94}. All pixels in the map that are connected with each other and above $A_\mathrm{V}^\mathrm{tail}$ are considered as one clump. We emphasize that we do not make an effort to identify single-peaked structures nested inside the contour defined by $A_\mathrm{V}^\mathrm{tail}$, because we particularly want to examine the mean physical parameters inside regions defined by the $A_\mathrm{V}^\mathrm{tail}$ threshold. Therefore, there can be numerous distinct column density peaks (of any column density higher than $A_\mathrm{V}^\mathrm{tail}$) nested inside the clumps. In terms of the clumpfind2d algorithm, this approach equals to using $A_\mathrm{V}^\mathrm{tail}$ as a threshold level for structure detection, but not defining any additional column density levels that would be used in detecting peaks inside this parental structure.

  % Results of these experiments

\begin{figure*}
   \centering
      \includegraphics[bb = 45 10 470 335, clip=true, width=0.38\textwidth]{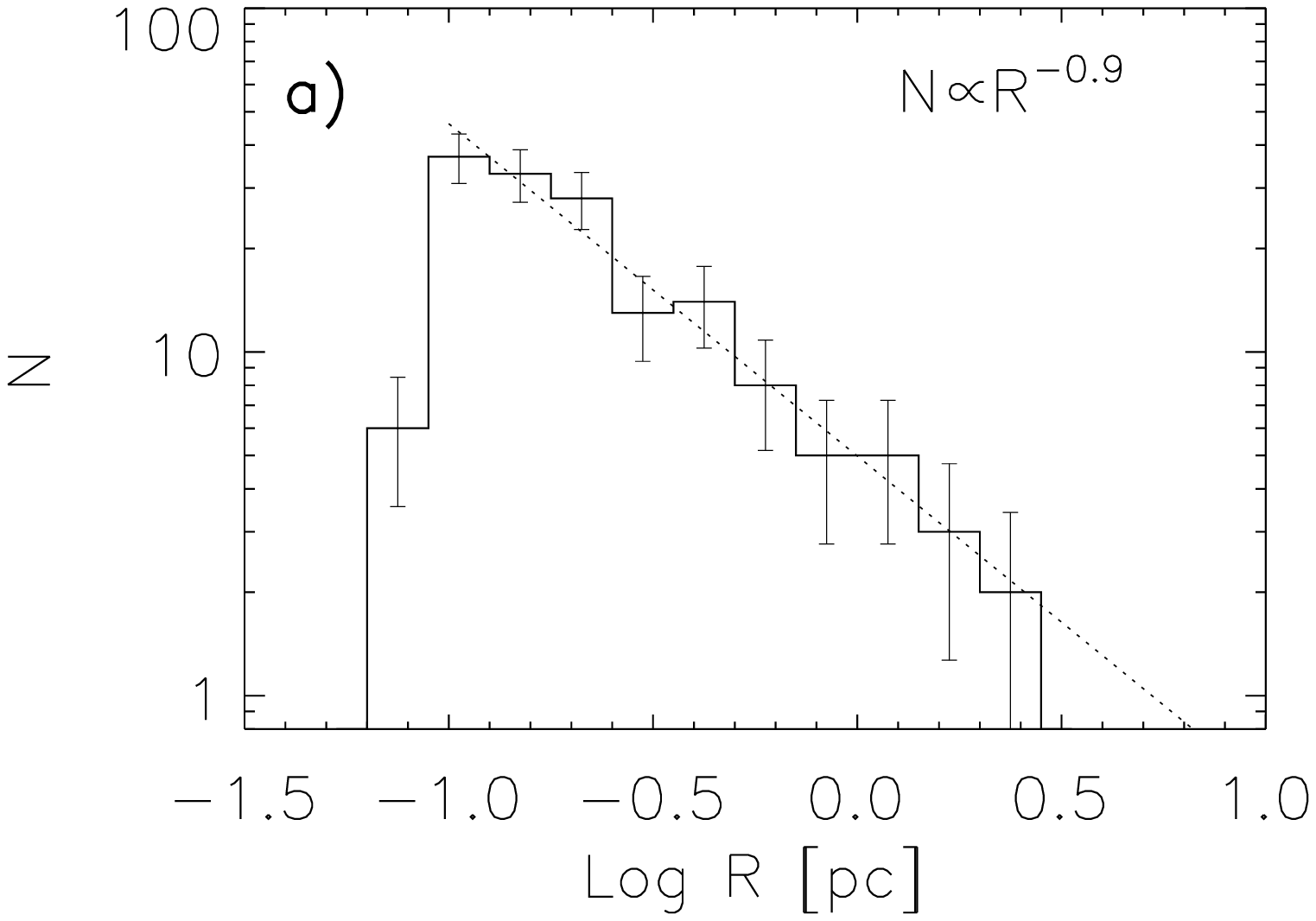}      \includegraphics[bb = 45 10 470 335, clip=true, width=0.38\textwidth]{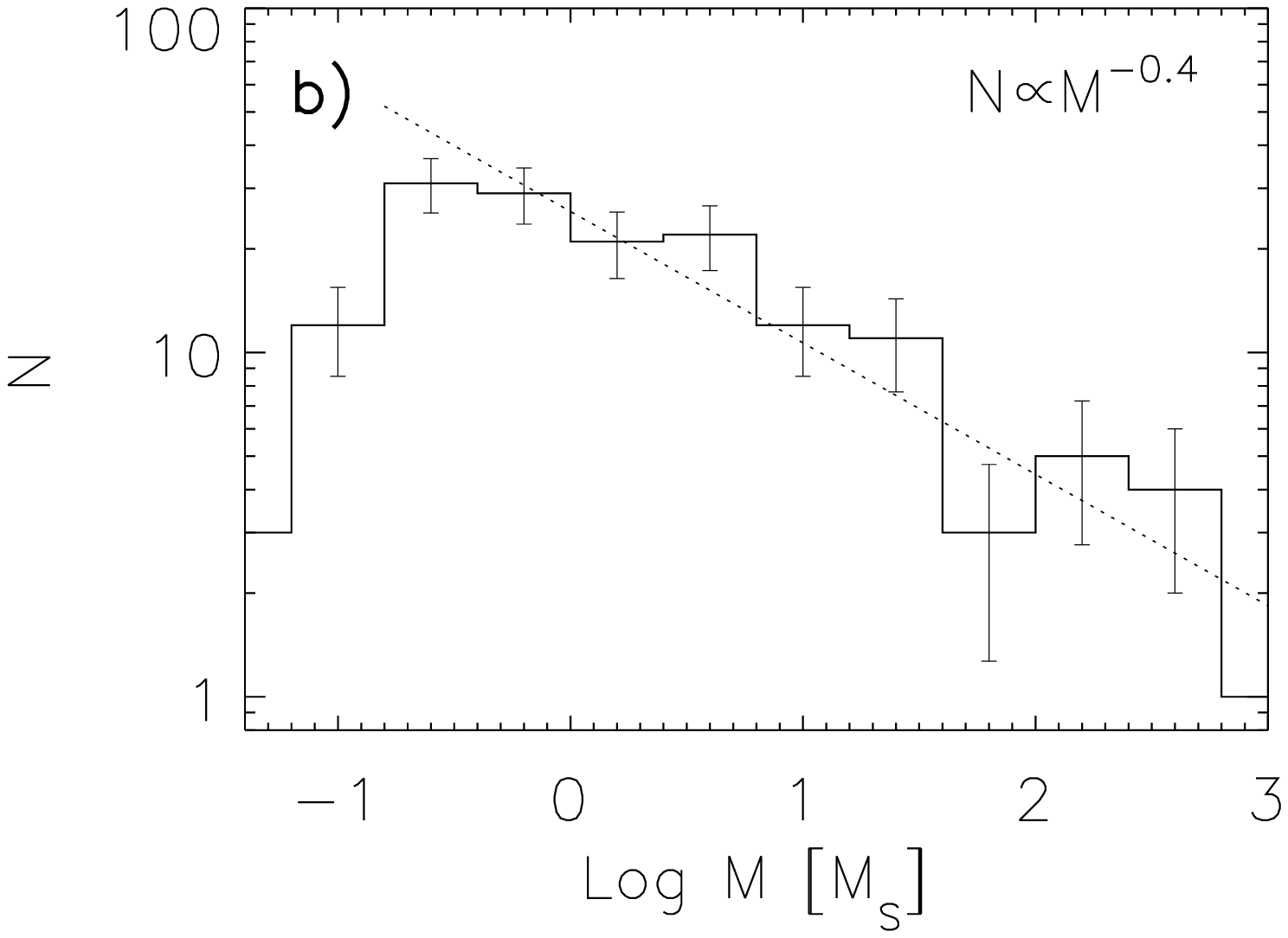}   
\includegraphics[bb = 5 10 500 335, clip=true,width=0.44\textwidth]{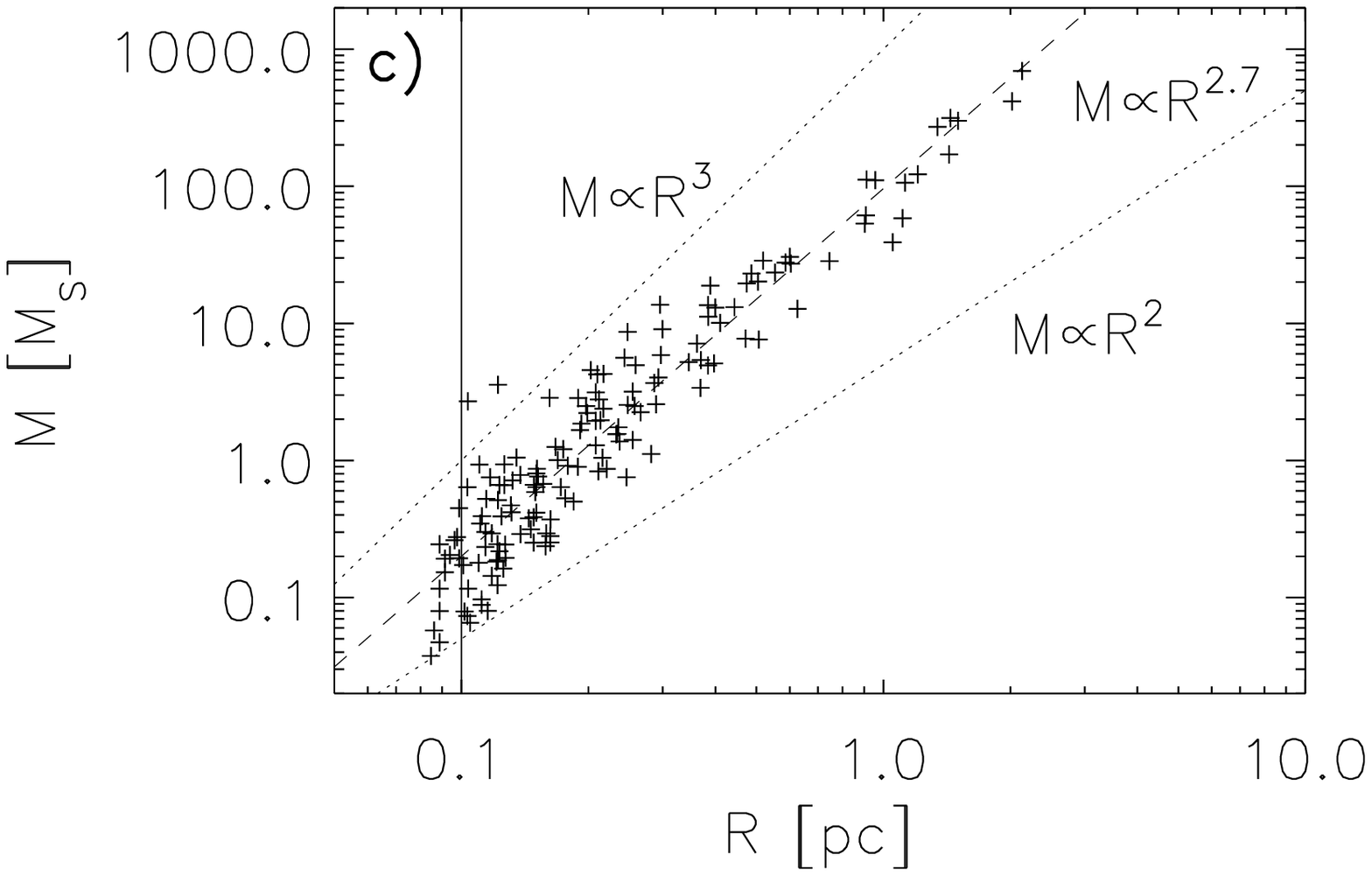}  \includegraphics[bb = 45 10 500 335, clip=true, width=0.4\textwidth]{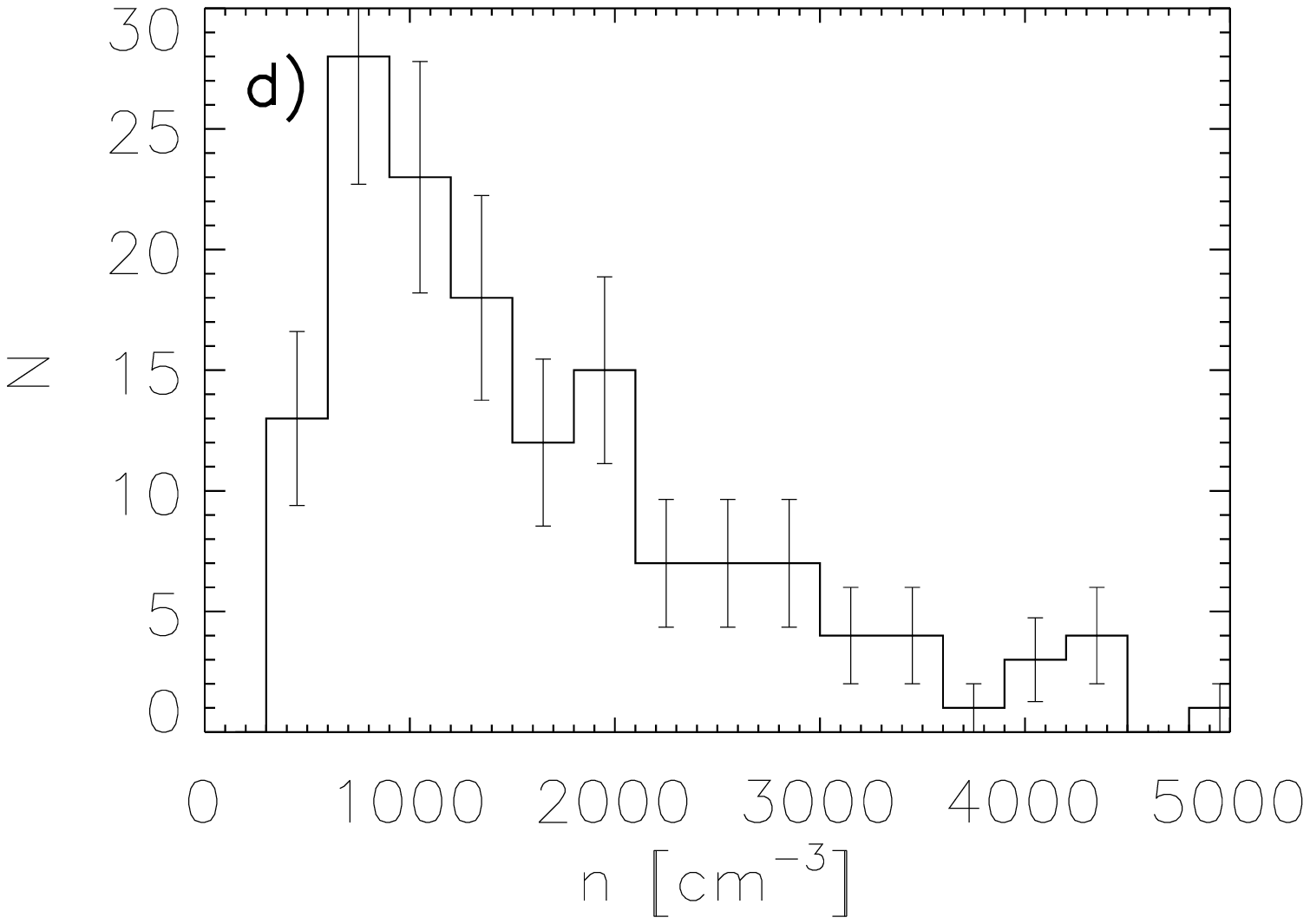}
      \caption{Characteristics of the structures (clumps) defined by a thresholding at $A_\mathrm{V}^\mathrm{tail}$ (see \S\ref{sec:mass} in text). {\bf a) }The size distribution. Error bars show the $\sqrt{n}$ uncertainty. The dotted line shows the least-squares fit to the distribution, with the slope $-0.9 \pm 0.2$. {\bf b) }The same for the mass distribution. The linear fit results in the slope $-0.4 \pm 0.2$. {\bf c) }The mass-radius relation. Overplotted are slopes indicating $M \propto R^3$ and $M \propto R^2$ (dotted lines), the linear fit to the data points which has a slope $2.7 \pm 0.2$ (dashed line), and the resolution limit $R = 0.1$ pc of the data (solid line). {\bf d) }The distribution of mean volume densities.
              }
         \label{fig:all_clouds}
   \end{figure*}

This exercise resulted in identification of $\sim 10$ clumps from each cloud complex that show wide ranges of sizes and masses. The effective radii of the clumps, defined as $R = \sqrt{A / \pi}$ where $A$ is the area, varies between $\sim 0.1 - 3$ pc. Figure \ref{fig:all_clouds} shows the size distribution of all clumps in all clouds, showing that smaller regions are always more numerous than larger ones down to the resolution limit of our data ($R\approx 0.1$ pc). The size distribution has a power-law-like shape with the approximate slope of $-0.9 \pm 0.2$. Figure \ref{fig:all_clouds} also shows the mass distribution of the clumps, calculated by integrating Eq. \ref{eq:mass_dense} over the clump. The mass distribution covers roughly four orders of magnitudes between $10^{-1}-10^3$ M$_\odot$, with approximately a power-law distribution that has a slope of $-0.4 \pm 0.2$. This slope is flatter than typically observed for the mass distributions of cores in the clouds \citep[$\sim -1.3$, e.g.][]{mot98, alv07, and10}, being closer to the slopes derived for molecular clouds or CO clumps within individual clouds \citep[$\sim -0.6$, e.g.][]{wil95, kra98, bli07}. It is, however, possible that the derived slope is affected by the blending of clumps with each other. Such blending can make the detection efficiency of clumps a function of the clump mass, and thereby affect the slope of the observed mass function \citep[e.g.,][]{kai09pipe, pin09}. The mass-to-size relation for the clumps, also shown in Fig. \ref{fig:all_clouds}, follows approximately the relation $M \propto R^{2.7 \pm 0.2}$, again being close to what has been derived for CO clumps within clouds \citep[e.g.,][]{car87}. The relation is also in agreement with what is expected for constant mean volume density spheres and steeper than expected for clouds in agreement with Larson's relations ($M \propto R^2$). Larson's mean-density size relationship, $\rho \propto R^{-1}$, however, has been questioned by \citet{bal02}, suggesting that it is an observational artifact due to limited dynamical range in column density. Our mass-radius relation, $M \propto R^{2.7\pm 0.2}$ can also be seen as an indicator of the fractal dimension of the cloud, $D\approx 2.7\pm 0.2$. This range is consistent with the largest values in the range, $D = 2.3 - 2.7$, previously found by \citet{elm96}, indicating fairly space-filling column density structures \citep{san05, fed09}.

The mean volume densities in the clumps, defined as $\overline{n} = M / (4/3\pi R^3) / m_\mathrm{H}$, is on the order of $10^3$ cm$^{-3}$. The mean volume densities of all clumps in all clouds have a distribution that peaks strongly at $\overline{n} \approx 0.8 \times 10^3$ cm$^{-3}$ (shown in Fig. \ref{fig:all_clouds}). The peak of the distribution is also relatively narrow, with the mean density being between $\overline{n} = 0.4-2.1 \times 10^3$ cm$^{-3}$ for 90 \% of the clumps. We also calculated the mean density of the diffuse component in the clouds. This is straightforwardly defined for each cloud without the clumpfinding process, since the diffuse component is always a uniform structure. Similarly with the density of the clumps, the volume was calculated using the effective radius $R = \sqrt{A / \pi}$ for the cloud. The resulting mean densities, listed also in Table \ref{tab:clouds}, are $\overline{n}_\mathrm{diff} = 1-2 \times 10^2$ cm$^{-3}$, i.e. $5-10$ times smaller than the mean densities of the clumps identified from the dense component. 

%-----------------------------------------------------------------------------------------
\subsubsection{Correlation with CO and linewidth}
%-----------------------------------------------------------------------------------------
\label{sec:velocity}

% Comparison of the spatial extent

\begin{figure*}
   \centering
      \includegraphics[width=0.9\textwidth]{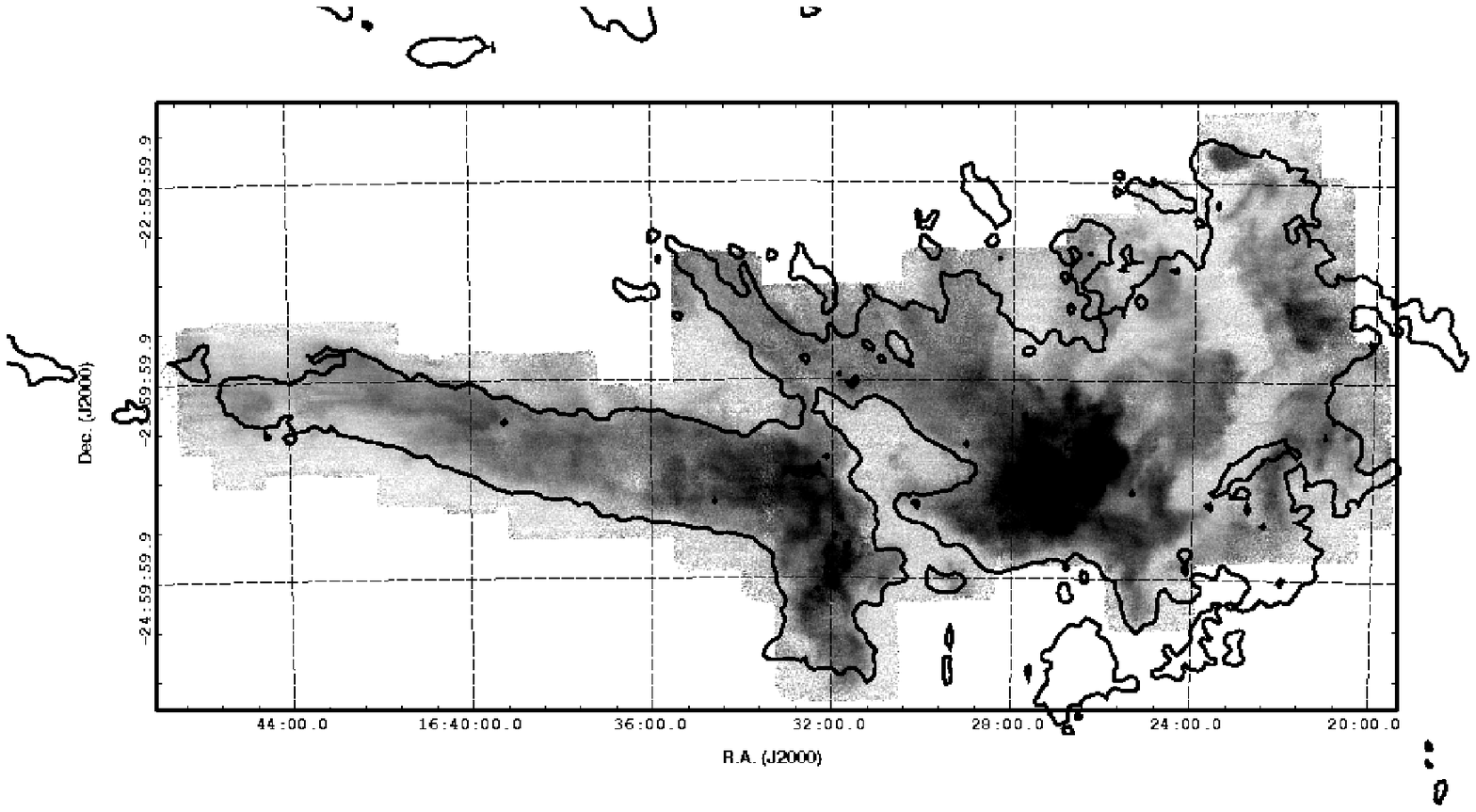}      
      \includegraphics[width=0.95\textwidth]{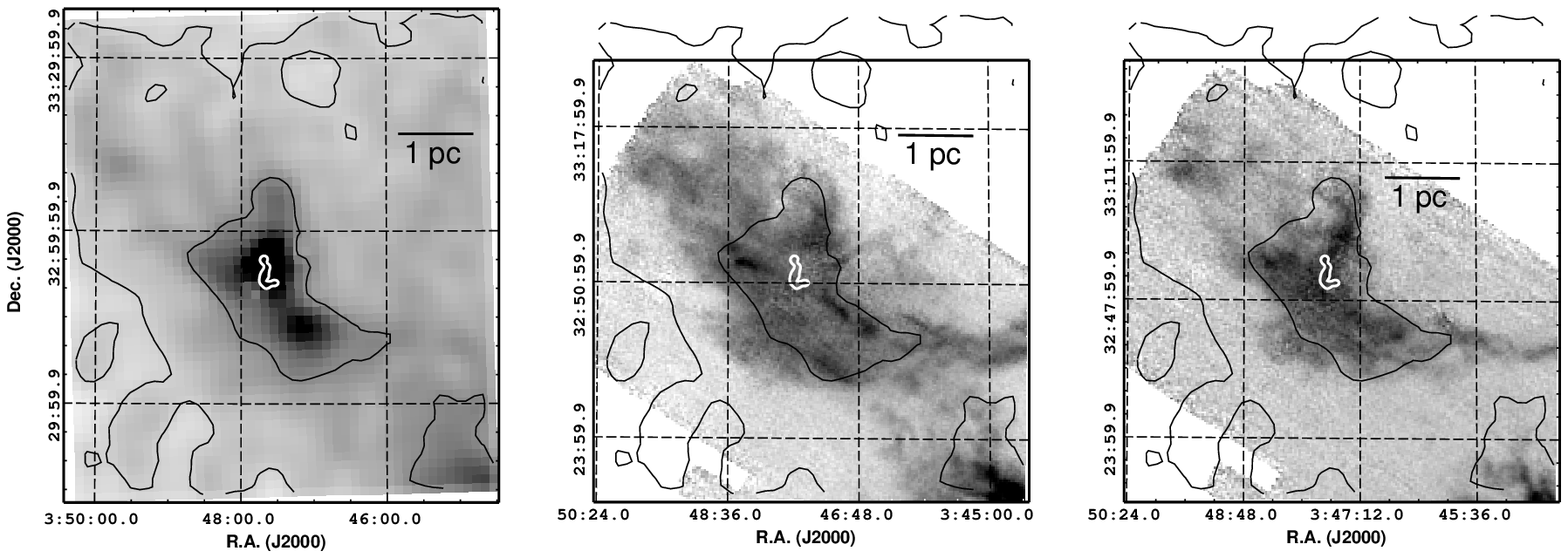}
      \caption{Comparison of the dense component, i.e. structures above $A_\mathrm{V} > A_\mathrm{V}^\mathrm{tail}$, with the CO molecular line emission. {\bf Top: } $^{12}$CO line emission from the COMPLETE survey for Ophiuchus, with a contour of $A_\mathrm{V}^\mathrm{tail} = 2.8$ mag overplotted. {\bf Bottom row: }Similar comparison for the B5 globule in Perseus. The left panel shows the extinction map, with black contours at $A_\mathrm{V} = [1, 3]$ mag. The white contour shows the extent of the coherent core in which the linewidth of the NH$_3$ molecule emission drops abruptly, identified by \citet{pin10jaime}. The center and right panels show the same for the $^{12}$CO and $^{13}$CO line emission, respectively. 
              }
         \label{fig:oph_co}
   \end{figure*}

In Fig. \ref{fig:oph_co}, we demonstrate how the spatial extent of the dense component compares to the common molecular line tracer observations. We use as an example the Ophiuchus and Perseus clouds for which large-scale $^{12}$CO and $^{13}$CO data are publicly available through the COMPLETE survey \citep{rid06a}. Figure \ref{fig:oph_co} shows the $^{12}$CO total antenna temperature map of the Ophiuchus cloud with a contour of $A_\mathrm{V}^\mathrm{tail} = 2.8$ mag overplotted. In Ophiuchus, thresholding at $A_\mathrm{V}^\mathrm{tail}$ separates two larger clumps: the main cluster region, and the streamers leading east from the cluster (there are additional clumps identified outside the coverage of the CO emission). In general, the $A_\mathrm{V}^\mathrm{tail}$ contour coincides quite well with the extent of the $^{12}$CO line emission data (1-$\sigma$ rms error of CO data is 0.98 K), while $^{13}$CO is spatially less extended. Figure \ref{fig:oph_co} also shows a similar comparison on a smaller spatial scale for the B5 globule in the Perseus cloud.

% Virial parameters

\begin{figure}
   \centering
   \includegraphics[width=\columnwidth]{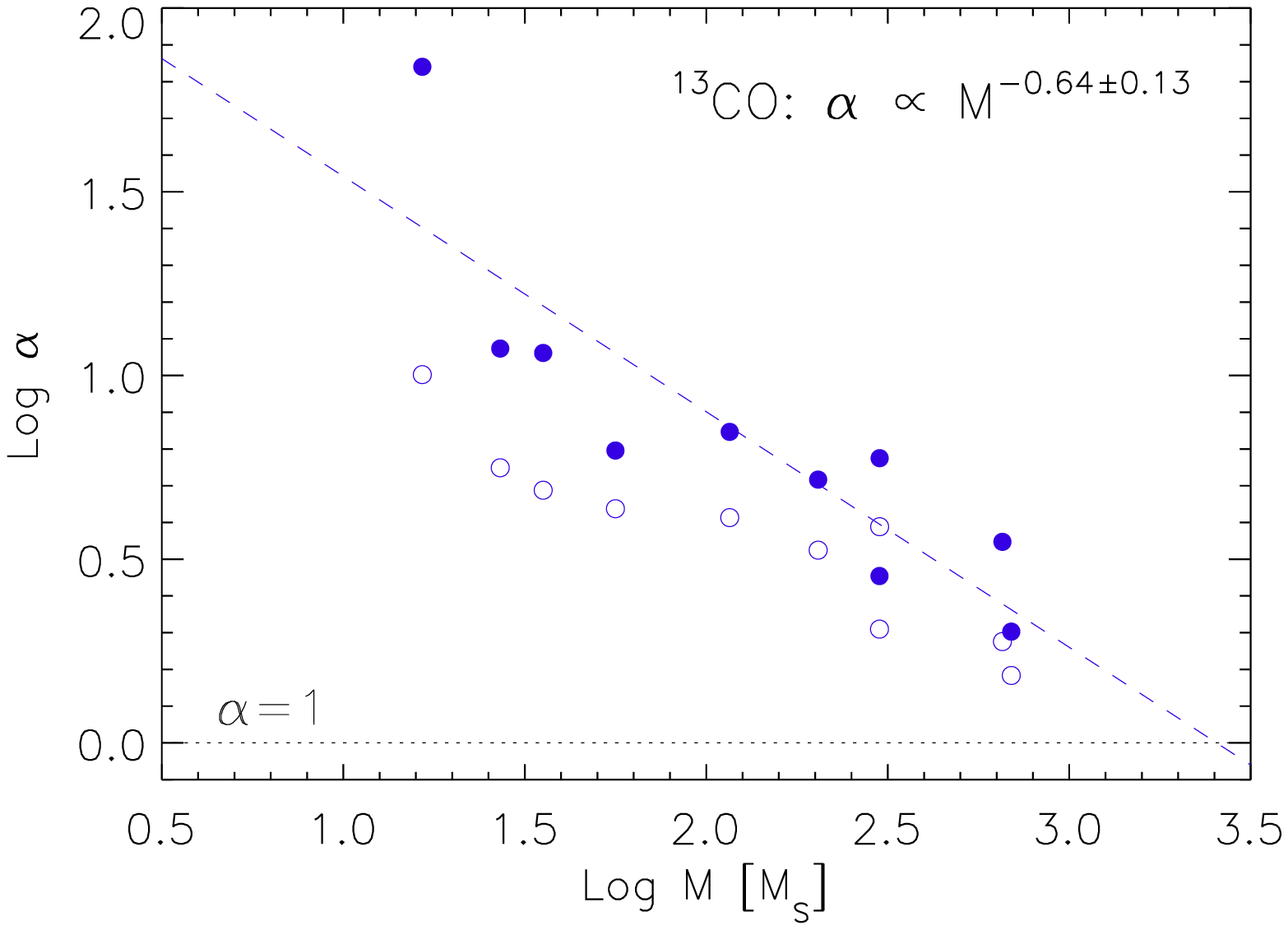}
   \includegraphics[width=\columnwidth]{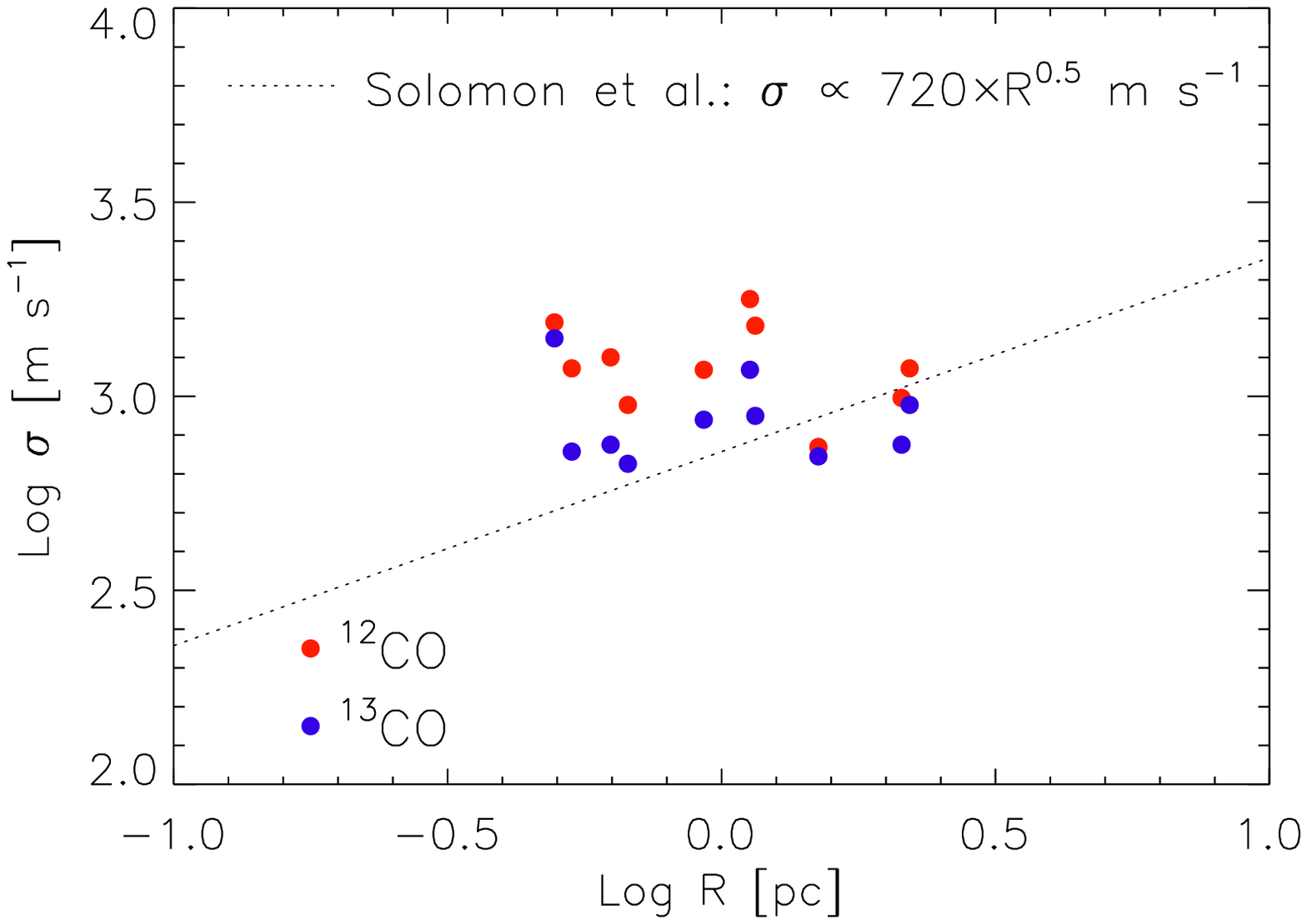}
         \caption{{\bf Top: }Virial parameters derived for clumps identified in the Ophiuchus and Perseus clouds using thresholding at $A_\mathrm{V} = A_\mathrm{V}^\mathrm{tail}$. In calculating the virial parameter, linewidths derived from $^{13}$CO data from the COMPLETE survey \citep{rid06a} were used. The dashed line shows the slope corresponding to the relation $\alpha \propto M^{-2/3}$, predicted for clumps confined by external pressure \citep{ber92}. A linear fit to the data points yields the slope $-0.64 \pm 0.13$. The open circles show the modified virial parameters of the same clumps, $\widetilde{\alpha}$, that take into account the external confining pressure (see \S\ref{sec:balance} in text). The dotted line indicates the $\alpha=1$ (and $\widetilde{\alpha}$ = 1) level. {\bf Bottom: }Size-linewidth relation for the same clumps, shown for both $^{12}$CO (red) and $^{13}$CO (blue). The dotted line shows the $\sigma \propto R^{0.5}$ relation \citep{sol87}, not a fit to the data points.}
         \label{fig:alpha}	
   \end{figure}

We identified 10 clumps in the Ophiuchus and Perseus clouds that are fully within the region covered by the COMPLETE survey. We estimated the virial parameters of these clumps, defined as the ratio of kinetic-to-gravitational energies in the clump \citep{ber92}: 
\begin{equation}
\alpha = \frac{5 \sigma^2 R}{G M}, 
\label{eq:a_vir}
\end{equation}
where $G = 1/232$ $\mathrm{M}_\odot^{-1} \  \mathrm{pc} \  ($km s$^{-1})^{2}$ is the gravitational constant and $\sigma$ the velocity dispersion. The linewidths were estimated from both the $^{12}$CO and $^{13}$CO data by calculating the mean spectrum over the clump and making a simple gaussian fit to the peak of it. The mass was calculated from the extinction data following Eq. \ref{eq:mass_dense}. This calculation yielded virial parameters $\alpha = 3-100$ for the clumps. The virial parameters correlate with the mass of the clumps approximately in a power-law fashion (Fig. \ref{fig:alpha}). A simple linear least-squares fit to the data points yields the slopes of $-0.69 \pm 0.12$ and $-0.64 \pm 0.13$ for $^{12}$CO and $^{13}$CO, respectively. This relation is consistent with the prediction for clumps confined by ambient pressure from the medium surrounding them \citep{ber92} and such relation has been previously observed for clumps identified from CO emission data \citep[e.g.,][]{ber92, wil95, lad08}. It is, however, clear that the determination of the virial parameters suffers from likely non-gaussian errors, arising most pressingly from the uncertainty in determining the linewidth that would well trace most of the gaseous material in the cloud. Therefore, we consider this observed correlation indicative, although clearly not well constrained. It is, however, evident that having $\alpha >> 1$, these clumps are not gravitationally bound entities (although they can be significantly supported by other forces, as will be discussed later). This is unsurprising, as CO clumps in molecular clouds are generally observed to have high virial parameters \citep[e.g.,][]{car87, ber92, fal92}.

Figure \ref{fig:alpha} also shows the size-linewidth relation for the clumps in Ophiuchus and Perseus. The data are scattered with no clear correlation, although similarly with the virial parameter - mass relation determining the correlation is hampered by the small number of clumps. We note that if the size and linewidth are uncorrelated, the relation between the virial parameter and mass established above ($\alpha \propto M^{-2/3}$) implies that $M \propto R^3$, i.e. the mean volume densities of the clumps are constant (see Eq. \ref{eq:a_vir}). Indeed, this is in agreement with the characteristics of the clumps derived in \S\ref{sec:size}, i.e. that the mass-radius relation follows approximately $M \propto R^{2.7 \pm 0.1}$, and that the mean volume densities of the clumps strongly peak around a characteristic value of $\sim 10^3$ cm$^{-3}$.

%[@@@ THIS IS THE SCALE WHERE CO DROPS, see McKee \& Ostriker 2007, Pineda et al. 2010, etc.]

%[@@@ Heyer, Carpenter \& Snell 2001]

%*************************************************************
\subsection{Pressure confinement of molecular clumps}
%*************************************************************
\label{sec:balance}

As demonstrated in \S\ref{sec:mass}, the mass in the clumps above $A_\mathrm{V}^\mathrm{tail}$, i.e. in the dense component, accounts only for the minor fraction of the total gaseous mass of a cloud. In other words, the clumps are surrounded by a medium whose total mass (and spatial extent) greatly exceeds that of their own. As an example, the most massive clump in our cloud sample has the mass of about 10 \% of the mass of the whole cloud (for other clumps, the fraction is much smaller). Likewise, as shown in \S\ref{sec:size}, the mean density of the clumps is close to an order of magnitude higher than the mean density of the medium surrounding them. Thus, it seems reasonable to consider the gravitational force of the surrounding medium as a source of external pressure supporting the clumps. In the following, we follow the formulation of \citet{ber92} and examine the scale of external support provided to the clumps by the diffuse medium surrounding them.

The basic condition for the virial balance of a clump is:
\begin{equation}
0 = W + 2(T-T_\mathrm{ext}) + \cal{M}.
\label{eq:virialbalance}
\end{equation}
In this equation, $W$ is the potential energy:
\begin{equation}
W = -\frac{3}{5} \frac{GM^2}{R}, 
\end{equation}
and $T$ and $T_\mathrm{ext}$ are the kinetic energy of the clump and its surface term:
\begin{equation}
T = 3 / 2 \times P_\mathrm{kin} V,
\end{equation}
\begin{equation}
T_\mathrm{ext} = 3 / 2 \times P_\mathrm{ext} V.
\end{equation}
$\cal{M}$ is the magnetic energy which we neglect for simplicity. The virial balance equation expressed in terms of pressure is then: 
\begin{equation}
P_\mathrm{kin} =  P_\mathrm{gr} +P_ \mathrm{ext}  .
\label{eq:pressurebalance}
\end{equation}
In this, the total kinetic pressure of the clump, $P_\mathrm{kin}$, is the sum of both thermal and non-thermal components:
\begin{equation}
P_\mathrm{kin} = \rho (\sigma_\mathrm{T}^2 + \sigma_\mathrm{NT}^2), 
\label{eq:Pkin}
\end{equation}
and $P_\mathrm{gr}$ is the gravitational pressure of the clump supporting it against expansion:
\begin{equation}
P_\mathrm{gr} = -1/3 \times W / V = (\frac{4\pi}{15}) G \  (\overline{\rho}R)^2.
\label{eq:Pgr}
\end{equation}
$P_\mathrm{ext}$ is the pressure external to the clump. Under the assumption that molecular cloud complexes are close to gravitational virial equipartition \citep[e.g.,][]{lar81, hey01}, a supporting external pressure is directed to a clump. This pressure arises from the turbulent pressure that balances the cloud against its own gravity. Since the cloud, as a whole, is close to virial equipartition, the turbulent pressure amounts to the gravitational pressure of the cloud (analogously to Eq. \ref{eq:Pgr}), but we adopt a slightly modified expression that takes into account that the cloud is not spherical \citep{ber92}. With the definition of the mean mass surface density, $\overline{\Sigma} = M / (\pi R^2)$, the external pressure supporting clumps against dispersal is:
\begin{equation}
P_\mathrm{ext} = P_\mathrm{gr}^\mathrm{cloud} = \large(\frac{3\pi a_1}{20}\large)G\overline{\Sigma}^2 \phi_\mathrm{G}, 
\label{eq:ext_pressure_bertoldi}
\end{equation}
where $a_1$ and $\phi_\mathrm{G}$ are numerical constants related to cloud morphology whose value can be evaluated as prescribed in \citet{ber92}. As an example of the order-of-magnitude of these pressures, using the typical $^{13}$CO linewidth of $\sigma = 0.75$ km s$^{-1}$ for a $R=1$ pc sized clump, $\overline{n}_\mathrm{diff} = 150$ cm$^{-3}$, $\overline{n}_\mathrm{clump} = 800$ cm$^{-3}$, and the mean mass surface density $\overline{A}_\mathrm{V} = 2$ mag yields the pressure ratios $P_\mathrm{kin} \approx 10 \times P_\mathrm{gr} \approx  4 \times P_\mathrm{ext}$. In other words, the pressures supporting the clumps against dispersal amount in total to about one third of the pressure driving their dispersal.

In the following, we examine these pressures for a population of clumps whose properties equal to those derived for the clumps identified in this paper. In \S\ref{sec:velocity}, we showed that the observed virial parameters of the clumps scale with their masses \citep[which is predicted for clumps confined by external pressure, ][]{ber92}:
\begin{equation}
\alpha = \frac{2T}{|W|} = \frac{P_\mathrm{kin}^\mathrm{clump}}{P_\mathrm{gr}^\mathrm{clump}} = c_1 \times M^{-2/3},
\label{eq:alpha}
\end{equation}
where $c_1$ is a proportionality constant. It directly follows from this dependence that the internal kinetic pressure of the clumps is only a function of their density (Eqs. \ref{eq:alpha} and \ref{eq:Pgr}), and thereby the kinetic pressure is constant for a population of constant density clumps. The ratio of outwards to inwards pressures for a clump is then:
\begin{equation}
\frac{P_\mathrm{out}^\mathrm{clump}}{P_\mathrm{in}^\mathrm{clump}} = \frac{P_\mathrm{kin}}{P_\mathrm{ext}+P_\mathrm{gr}} = \frac{c_1 (\frac{4\pi}{3})^{-2/3}}{(\overline{\rho})^{2/3}R^2 + \frac{9a_1\phi_\mathrm{G}}{16}\frac{(\overline{\Sigma})^2}{(\overline{\rho})^{4/3}}}.
\label{eq:ratio}
\end{equation}

\begin{figure*}
   \centering
   \includegraphics[bb=38 10 480 335, clip=true, width=0.7\columnwidth]{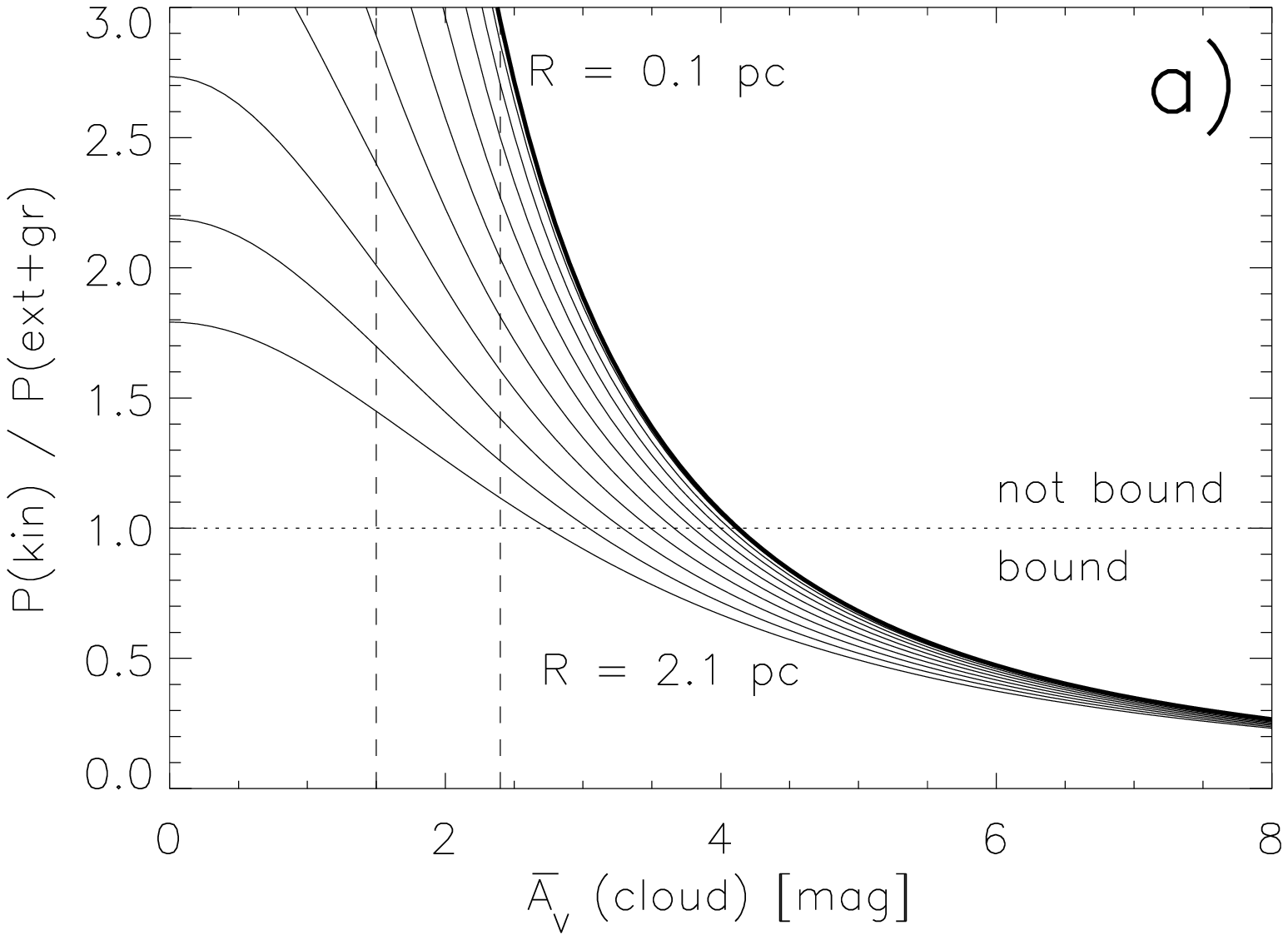}   %\includegraphics[bb=38 10 480 335, clip=true, width=0.664\columnwidth]{balance2.eps}   
   \includegraphics[bb=38 10 480 335, clip=true, width=0.7\columnwidth]{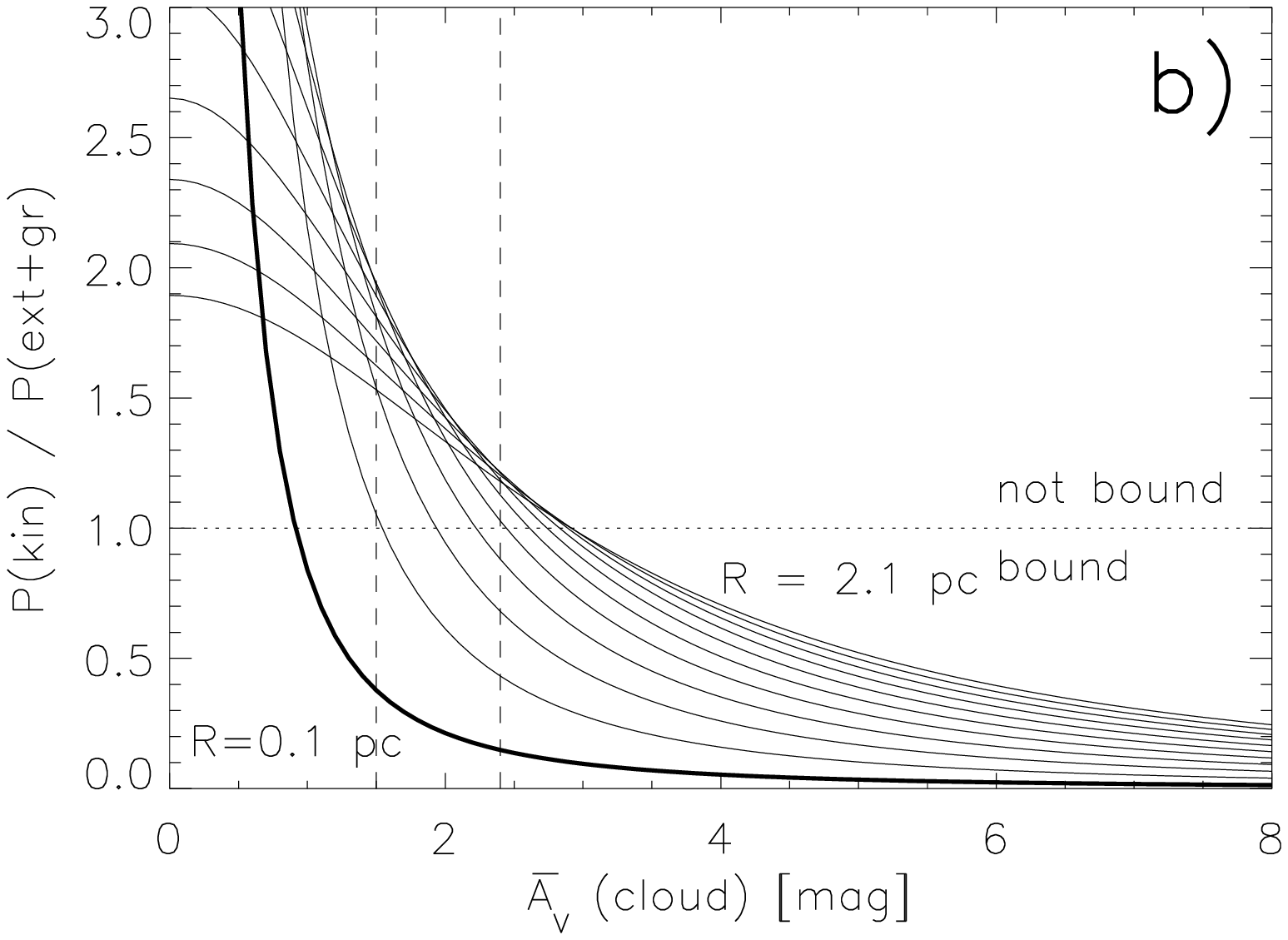}
      \caption{{\bf a) }The ratio of the pressures supporting a clump against a collapse (total kinetic pressure) to the pressures promoting it (external and gravitational pressure) in clumps with $\overline{n} = 800$ cm$^{-3}$. The different curves are for clumps with radii $R = 0.1-2.1$ pc, in steps of 0.2 pc. The dashed vertical lines indicate the interval of the observed mean extinctions (see Table \ref{tab:clouds}). {\bf b) }The same, but using Larson's size-linewidth relation to calculate the kinetic energies of the clumps (i.e., Eq. \ref{eq:alpha_larson} instead of Eq. \ref{eq:alpha}).}
         \label{fig:balance}
   \end{figure*}

Figure \ref{fig:balance}a illustrates this ratio (Eq. \ref{eq:ratio}) as a function of the mean mass surface density $\overline{\Sigma}$ (in units of $A_\mathrm{V}$) for clumps of different sizes ($R = 0.1-2.1$ pc). As the mean density, we used the value $\overline{n} = 800$ cm$^{-3}$ shown earlier to be the peak of the mean densities in the clumps (\S\ref{sec:size}). Figure \ref{fig:balance} shows that the transition from a regime where structures are unbound to a regime where they are bound occurs around $\overline{A}_\mathrm{V} \approx 4$ mag ($\overline{\Sigma} \approx 80$ M$_\odot \mathrm{\ pc}^{-2}$) for clumps over a wide range of sizes  ($R = 0.1-2.1$ pc). This value is larger than the observed mean mass surface densities by a factor of about $\sim 2$ (see Table \ref{tab:clouds}), and therefore, the external pressure is lower than the internal kinetic energy of the clump by a factor of $\sim 4$.

We have so far assumed that the kinetic energies of the clumps scale with mass as shown by Eq. \ref{eq:alpha}. In principle, this scaling is supported by the linewidth data of clumps showing rather constant linewidths (Fig. \ref{fig:alpha}). This observation is, however, hampered by the poor statistics we could achieve with the available data. Therefore, we consider also a case where the kinetic energies of the clumps scale according to a Larson-like size-linewidth relation \citep{sol87}:
\begin{equation}
\alpha^\mathrm{clump} = \frac{2T}{|W|} = \frac{P_\mathrm{kin}^\mathrm{clump}}{P_\mathrm{gr}^\mathrm{clump}} = \frac{5\times 0.72 (\frac{R}{1 \ \mathrm{pc}} )^{0.5}R}{GM}.
\label{eq:alpha_larson}
\end{equation}
Again, Fig. \ref{fig:balance}b shows the ratio of outwards to inwards pressures for clumps (Eq. \ref{eq:ratio}) as a function of mean mass surface density of the cloud. Using this scaling of kinetic energies, the balance occurs approximately at the level of the observed mean mass surface density ($\overline{A}_\mathrm{V} \approx 2$ mag). While our observations seem to favor kinetic energy scaling with mass (see also discussion in \S\ref{sec:discussion}), this example illustrates the behavior of the pressure balance in another plausible scaling scheme, suggesting that the external pressure indeed can be close to the internal kinetic energy of the clumps.

Finally, we illustrate the net effect of the pressures to the individual observed clumps by calculating modified virial parameters for the clumps for which we have CO data. In this modified virial parameter, we take into account the pressure external to the clumps:
\begin{equation}
\widetilde{\alpha}^\mathrm{\ clump} = \frac{P_\mathrm{kin}^\mathrm{\ clump}}{P_\mathrm{gr}^\mathrm{\ clump}+P_\mathrm{ext}}.
\label{eq:a_tilde}
\end{equation}
The modified virial parameters are shown in Fig. \ref{fig:alpha} together with the traditional virial parameters that include only the gravitational and kinetic energies of the clumps. In agreement with the earlier results, the modified virial parameters are clearly smaller compared to the ones resulting from Eq. \ref{eq:a_vir}, but still somewhat larger than unity.

%**************************
%**************************
\section{Discussion}  %*
%**************************
%**************************
\label{sec:discussion}

\subsection{Pressure confinement of the clumps}

% Starting paragraph

We described in the previous section a new approach to characterize structures observed in molecular clouds. In particular, we used the observed gas column density PDFs of molecular clouds to define a population of clumpy structures (dense component) embedded in the extended, interclump medium (diffuse component). The transition between these components occurs at the extinction threshold $A_\mathrm{V}^\mathrm{tail} = 2-4$ mag, or equivalently, at $\Sigma^\mathrm{tail} = 40-80$ M$_\odot$ pc$^{-2} = 0.008-0.017$ g cm$^{-2}$. This level is relatively constant, being in the quoted range in every cloud except one (Serpens, cf. Table \ref{tab:clouds}) for which it could be reliably determined. The dense component becomes dominant at $A_\mathrm{V} = 3-8$ mag. The mass of the dense component is between $1-20$ \% of the total mass of the cloud which we defined as the total mass above $A_\mathrm{V} > 1$ mag. The clumps of the dense component show roughly power-law-like distributions of sizes and masses, covering wide dynamical ranges (see Fig. \ref{fig:all_clouds}). However, the components are characterized by remarkably constant mean volume densities of $\overline{n} \approx 10^3$ cm$^{-3}$ and $\overline{n} \approx 1.5 \times 10^2$ cm$^{-3}$ for the dense and diffuse components, respectively. 

% Clumps identified using PDFs resemble 13CO clumps

The clumps identified using the column density PDFs of the clouds in this study are very similar to the \emph{gravitationally unbound} $^{13}$CO clumps identified in several studies in the past \citep[e.g.][]{car87, ber92, wil95, lad08}. In particular, the mean densities, and mass-radius and virial parameter-mass relations derived in \S\ref{sec:size}-\ref{sec:velocity} are in agreement to what has been derived for such CO clumps. Similarly, those studies have concluded that the external pressure can be a significant confining source for such clumps. A simple qualitative comparison between the extent of the regions selected with $A_\mathrm{V}^\mathrm{tail}$ and $^{13}$CO emission indeed suggests these methods may trace quite similar components in the clouds (see Fig. \ref{fig:oph_co}).

% In particular, it does not correspond to "transition to coherence"

It has been earlier suggested \citep{goo98, cas02}, and more recently directly observed \citep{pin10jaime} that there appears to be a sharp transition to dynamically coherent objects, or cores, at the scale where non-thermal motions cascade from the supersonic to subsonic regime \citep[the sonic scale, e.g.][]{vaz03, fed10}. In particular, \citet{pin10jaime} detected such transition in the B5 globule in the Perseus cloud, at the length scale on the order of $R \approx 0.1$ pc, in agreement with the sonic scale (and Larson's size-linewidth relation). The spatial extent of this coherent core is illustrated in Fig. \ref{fig:oph_co}, together with the extent of the clump defined by $A_\mathrm{V}^\mathrm{tail}$. The transition to coherence occurs clearly in a different column density regime than the break in the column density PDF defined by $A_\mathrm{V}^\mathrm{tail}$. Importantly, while the structural transition from supersonic- to subsonic velocities in cloud structure seems to be linked to a particular size-scale, \emph{the structural transition described by the $A_\mathrm{V}^\mathrm{tail}$ threshold is not size-dependent}. As shown in Fig. \ref{fig:all_clouds}, structures above $A_\mathrm{V}^\mathrm{tail}$ cover a large size- and mass range, with their number decaying in a roughly power-law-like fashion. Thus, the $A_\mathrm{V}^\mathrm{tail}$ threshold appears to be unrelated to the transition to coherent cores in the velocity structure. Unfortunately, the column density maps used in our work do not provide high enough spatial resolution to properly sample the PDF at the length scale of the transition to coherence. Therefore, we could not directly look for possible features the transition would induce to the PDFs.

% The pressure balance

We showed in Section \S\ref{sec:balance} that a significant external pressure from the surrounding cloud is imposed to the clumps we identified using the $A_\mathrm{V}^\mathrm{tail}$ threshold. This is especially the case if, instead of a constant kinetic pressure, clumps follow a Larson-like size-linewidth scaling relation. This indicates that the clumps may even be close to a pressure balance with their surroundings. The CO linewidths and virial parameters we derived for a small sample of clumps partially support this picture: virial parameters (as defined by Eq. \ref{eq:a_vir}) correlate with clump masses as predicted for pressure confined clumps, and the linewidths do not show clear correlation with clump sizes (we observe a nearly constant linewidth for the clumps). On the other hand, the modified virial parameters taking the external pressure into account (Fig. \ref{fig:alpha} and Eq. \ref{eq:a_tilde}) were somewhat in excess to unity for those clumps, implying that they may be over-pressurized and thus either an additional pressure component may be significantly affecting them, or they may be expanding. This result is, again, similar to what has been derived for $^{13}$CO clumps \citep[e.g.,][]{car87}.

\begin{figure}
   \centering 
    \includegraphics[width=\columnwidth]{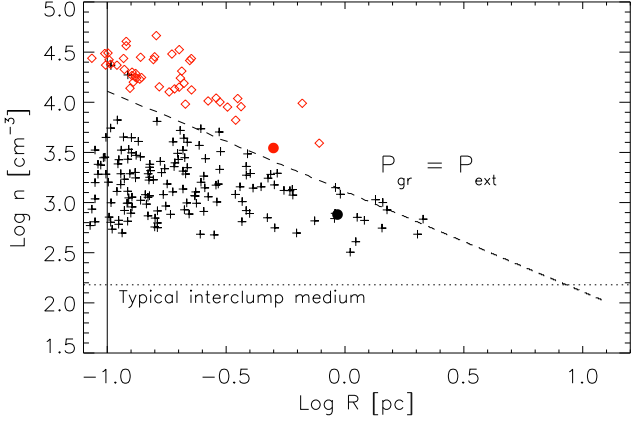}
      \caption{Relation between the mean density and size of the structures (clumps) defined by thresholding at $A_\mathrm{V}^\mathrm{tail}$. The dotted line shows the typical density of the interclump medium (i.e. the density of the diffuse component). The dashed line shows the density above which the gravitational energy of the clump becomes larger than the external surface pressure ($P_\mathrm{gr} = P_\mathrm{ext}$). The red diamonds show the structures identified from the column density maps by thresholding at $A_\mathrm{V} = 3 \times A_\mathrm{V}^\mathrm{tail}$. The resolution limit of the column density data is at $\log{R} = -1.0$. The black solid circle marks the clump B5 in Perseus, and similarly, the red solid circle marks the same clump when using the threshold of $A_\mathrm{V} = 3 \times A_\mathrm{V}^\mathrm{tail}$.
              }
         \label{fig:all_clouds2}
   \end{figure}

The role of the internal gravitational pressure of the clumps is further illustrated in Fig. \ref{fig:all_clouds2}, which shows the mean density for all identified clumps as a function of the clump size (plus signs in the figure). A constant ratio of gravitational-to-external pressures defines a linear relationship in this plot with a slope of $-1$ (cf. Eq. \ref{eq:Pgr} and \ref{eq:ext_pressure_bertoldi}). The relation corresponding to $P_\mathrm{gr} = P_\mathrm{ext}$ is overplotted in Fig. \ref{fig:all_clouds2}. All clumps identified from the dense component have densities lower than this relation, implying that the gravitational energy is indeed small compared to the external pressure. The typical mean density $n = 150$ cm$^{-3}$ of the diffuse component is also shown, which obviously is clearly below the mean densities of the clumps.

Despite the limits imposed by the spatial resolution and dynamical range of the extinction maps, we can still examine the  role of gravity in the structures nested inside the clumps (i.e, in smaller-scale structures inside what we have defined as a clump). We show in Fig. \ref{fig:all_clouds2} with red diamonds a population of structures identified by an experiment in which we defined clumps using a threshold level $A_\mathrm{V} = 3 \times A_\mathrm{V}^\mathrm{tail}$. While this threshold is typically $A_\mathrm{V} = 6-12$ mag, such selection likely represents a population connected to star-forming regions, or at least, pre-stellar objects. The structures identified with this experiment are above the $P_\mathrm{gr} = P_\mathrm{ext}$ line\footnote{We note that this relation is not in any dependence to the pressure balance or virial status of the clumps (see Eq. \ref{eq:Pgr} and \ref{eq:ext_pressure_bertoldi}). In order to estimate the kinetic pressure of the clumps identified in this experiment, a tracer probing densities of these objects would have to be used instead of $^{12}$CO or $^{13}$CO.}. This demonstrates how gravitation becomes an increasingly important confining force for density enhancements nested inside the clumps. To illustrate one case where gravitation is known to eventually become the dominant force, we have marked into the diagram the star-forming clump B5 in Perseus (see Fig. \ref{fig:oph_co}). The clump defined by thresholding at $A_\mathrm{V}^\mathrm{tail}$ is marked with black filled circle, and the structure identified inside it with the threshold at $3 \times A_\mathrm{V}^\mathrm{tail}$ is marked with red filled circle. 

Given these results, we suggest that the observed organization of structures, identified with a new approach using the column density PDF, can be understood as a population of clumps significantly supported by the external pressure. This external pressure originates from the turbulent pressure outside the clumps, and in this framework, it is a consequence from the assumption that clouds as a whole are close to virial equipartition. Then, the break observed in the column density PDFs at $A_\mathrm{V}^\mathrm{tail}$ represents a transition from a diffuse inter-clump medium to clumps significantly supported by external pressure. This interpretation has some profound implications. Most pressingly, it implies that the external pressure from the large-scale cloud has an important role in the formation of molecular cloud structures over wide size and mass scales. This result is analogous to the recent work of \citet{lad08}, who found the external pressure a significant force in confining small-scale cores, or globules, in the Pipe Nebula. We note that in Paper I, we speculated that the transition in the PDFs could be related to a transition from a regime where gravitation is negligible to a regime where it is dominant. The results of this paper, however, revise this picture and more quantitatively connect the break to the pressure conditions of density enhancements with their surroundings.

We emphasize three main assumptions of this framework. First, we assume that molecular clouds are in virial equilibrium as a whole, in order to relate their gravitational pressure to the turbulent pressure confining the clumps. Although close to being in virial equilibrium, molecular clouds are also embedded in an external medium and may have been formed in large-scale converging flows, being highly dynamical objects, where surface terms play an important role \citep[see e.g.,][]{bal06, ban09, sch10}. Second, as can be seen from Fig. \ref{fig:balance}, the external pressure due to the turbulent pressure has a significant role in providing support for the clumps. The confining force exerted to a clump by it, however, depends on the isotropy of the turbulent flow surrounding the clump. Supersonic turbulence is highly anisotropic locally and will lead to transient formation and destruction of filaments and clumps. While estimating the net effect of the turbulent pressure in a more consistent way for each clump would require more detailed velocity information (and modeling of the cloud structure) than are available for this work, we estimated the average level the pressure may have on clumps implicitly in Eq. \ref{eq:ext_pressure_bertoldi}. Third, we neglected the magnetic field stress in our approach. The exact role of the magnetic fields in shaping cloud morphology is still under debate, but generally, it can provide confining pressure perpendicular to the field lines which supports clumps against both collapse and expansion. The magnitude of this support can be estimated roughly with $B^2/(8\pi)$, which generally leads to pressures similar to turbulent ram pressures for the field strengths of $B\approx 15\  \mu\mathrm{G}$ \citep{cru99}.  

Interestingly, \citet{cru10} recently found that at densities lower than $n \lesssim 300$ cm$^{-3}$ the magnetic field strength does not scale with density, implying that below that density the cloud material is likely channelled along the field lines. Above this threshold density, the field strength approximately scales with density as $B \propto n^{0.65}$. In the context of our work, the threshold density of $n \approx 300$ cm$^{-3}$ is in the regime between the diffuse component ($\overline{n}\approx 150$ cm$^{-3}$) and the clumps ($\overline{n}\approx 800$ cm$^{-3}$). This rises the interesting possibility that the break in the column density PDFs at $A_\mathrm{V}^\mathrm{tail}$ would be related to the change in the role of the magnetic field from dominant at low column densities to less significant at higher column densities. While \citet{cru10} suggests that this could be the regime where cloud structures become self-gravitating, our work rather suggests that this change in the $B$-$n$ relation could be due to a transition to structures that are not self-gravitating but confined due to their approximate pressure balance with their surroundings.

\subsection{Pressure confinement and star formation}

When coupled with the main result of Paper I, i.e. that non-star-forming clouds do not exhibit similarly strong tails (if any) in their PDFs as \emph{all} star-forming clouds do, the interpretation discussed in this paper leads to a picture in which the formation of pressure confined clumps occurs in clouds prior to (or at clearly higher rate than) the formation of gravitationally dominated cores. Indeed, in our sample of molecular clouds, pressure confined clumps are observed in some clouds that do not show active star formation (Musca, Cha III), or even high column density cores (Musca, Hacar et al. in prep.). This picture is further supported by the recent analysis of the stability of dense cores in the nearby, mostly quiescent Pipe Nebula \citep{lad08}. In this cloud, Lada et al. examined a sample of $\sim150$ cores of masses between $0.2-20$ M$_\odot$ ($R\approx 0.04-0.2$ pc). They concluded the core population in the Pipe to be pressure confined, gravitationally unbound entities. The cores in Lada et al. study were defined to be single-peaked (or at most a-few-peaked) entities, and thereby they may well represent the \emph{smallest scale} of the hierarchy whose \emph{largest scale} is represented by the structures identified in our study are. A similar result highlighting the role of external pressure in a star-forming cloud was recently published by \citet{mar10}. They investigated the stability of dense cores in the Ophiuchus cluster, in the regime of $R \lesssim 0.1$, and concluded that the external pressure has a significant role in the dynamics of the cores. These results clearly indicate that the hierarchy of structures nested inside clumps is affected by the external pressure all the way to the regime of dense star-forming cores. %Unfortunately, as shown in Paper I and discussed in \cite{lom06}, the column density PDF for the Pipe seems to be disturbed by a cloud physically unrelated to the Pipe that covers the main body of it, and therefore we could not further examine the possible hierarchy between these pressure confined cores and possible larger scale pressure confined clumps. Likewise, the column density regime probed by \citet{mar10} is not accessible via our extinction maps.

% Implications for star formation .

In the interpretation discussed above the formation of pressure bound clumps can be seen as a prerequisite for the formation of gravitationally bound cores. Given this, there evidently should be a relation between the occurrence of such clumps and star formation, even beyond the general observation that not all quiescent clouds show such clumps, while all star-forming clouds do. Therefore, it is interesting to consider the observed star-forming efficiencies and -rates in the clouds of our sample.

Recently, \cite{hei10} studied the star-forming activities of nearby molecular clouds as a function of the gas surface density (i.e, the Kennicutt-Schmidt law). In their work, Heiderman et al. used near-infrared extinction maps similar to those employed in this paper to derive gas surface densities. They examined the number of young stellar objects (YSOs) in the clouds identified using the Spitzer satellite data in different column density intervals and constructed the Kennicutt-Schmidt law for their cloud sample. In particular, they observed an abrupt drop in the star formation rate at $\Sigma \approx 50-100$ M$_\odot$ pc$^{-2}$ \citep[see Figs. 3 and 8 in][]{hei10}, leading them to suggest a threshold for star formation at $\Sigma_\mathrm{th} = 129 \pm 14$ M$_\odot$ pc$^{-2}$ ($A_\mathrm{V} = 8.6$ mag).
A very similar result was reached recently by \cite{lad10}, who examined the relation between the number of YSOs in nearby clouds and the amount of high column density material in them. They showed that the correlation between the mass of the gas and the number of YSOs identified in the clouds is strongest (i.e., the dispersion in the relation is smallest) at $A_\mathrm{K} \approx 0.8$ mag ($A_\mathrm{V} \gtrsim 7.3$ mag, or $\Sigma \approx 116$ M$_\odot$ pc$^{-1}$). We note that the dispersion of the SFR-surface density relation derived by \citet{lad10} starts to decrease already at surface densities lower than $\Sigma \approx 116$ M$_\odot$ pc$^{-1}$, reaching its minimum at that point.

The star formation thresholds derived in the studies above are slightly larger than the typical $A_\mathrm{V}^\mathrm{tail}$ values. However, we defined the $A_\mathrm{V}^\mathrm{tail}$ value as the point where the dense component, on average, becomes a significant excess over the diffuse component. Obviously, at such surface density the largest contribution to the PDF still comes from the underlying diffuse component, not from the excessive dense component. Typical surface density values at which the contribution of the tail to the PDF becomes dominant ($> 90$ \%) are around $A_\mathrm{V}^\mathrm{tail} (90 \%) \approx 3-8$ mag (listed in Table \ref{tab:clouds}). Such values would be very much in agreement with the threshold values derived by \citet{lad10} and \citet{hei10}, given the very different approaches used in these papers. Therefore, it seems plausible to interpret increase in star-forming activity to be related to the regime where the column density PDF is becoming completely dominated by the dense component.
 
In the context of clumps bound by external pressure, a natural threshold for star formation is introduced by the surface density at which pressure bound clumps form, which is around $A_\mathrm{V}^\mathrm{tail}$. Furthermore as discussed above, the mass above this threshold is in a direct connection to the SFR of the cloud, with the SFR increasing in a power-law manner with increasing gas surface density. \emph{This interpretation gives a physically motivated explanation for the star formation threshold occurring at relatively low surface densities and links it to an observed structural feature in the clouds.} Thus, we suggest a picture in which the formation of pressure bound clumps, and thereby the structural transition at $A_\mathrm{V}^\mathrm{tail}$, introduces a prerequisite for star formation, with the amount of mass in clouds above that limit directly proportional to the capability of the cloud to form stars.

%**************************
%**************************
\section{Conclusions} %*
%**************************
%**************************
\label{sec:conclusions}

In this paper, we presented an analysis of the large-scale, clumpy structures in nearby molecular clouds and of their stability. In particular, we described a new approach to identify structure in clouds using the observed column density PDFs. With this approach, we identified two distinctive components in them, referred to as the dense and diffuse components, and described their basic physical characteristics. We then examined the stability of the clumps in the dense component, especially by considering the scale of external pressure imposed to them by the medium surrounding them. The main conclusions of our work are as follows:

   \begin{enumerate}
   
      \item The transition between the diffuse and dense components occurs at a narrow range of column densities, $A_\mathrm{V}^\mathrm{tail} = 2-4$ mag, or $\Sigma^\mathrm{tail} = 40-80$ M$_\odot$ pc$^{-2}$. The dense component dominates the observed column density PDFs above $A_\mathrm{V} > 3-8$ mag. The total mass of the dense component is 1-20 \% of the total mass of the cloud, and thus always clearly smaller than the mass of the diffuse component. Clumps identified in the dense component show wide dynamical ranges of sizes ($0.1- 3$ pc) and masses ($10^{-1}-10^{3}$ M$_\odot$). However, the mean volume density of the clumps is remarkably constant, $\overline{n} \approx 10^3$ cm$^{-3}$. This is $\sim 5-10$ times larger than the mean volume density of the diffuse component, $\overline{n} \approx 1-2 \times 10^2$ cm$^{-3}$.

     \item The clumps identified using the column density PDFs are gravitationally unbound and the external pressure, caused by the turbulent pressure from the diffuse, large-scale cloud surrounding them, can provide significant support for them against dispersal. However, examination of the stability of a small sub-sample of clumps indicates that they may be over-pressurized and either expanding or additionally supported by a component not included in our analysis (e.g. magnetic field support). Then, the physical properties of the clumps resemble those of the clumps often identified from $^{13}$CO emission observations as structures of the lowest hierarchical level. 
     
     \item In \citet{kai09}, we showed that some non-star-forming clouds do not show the PDF break, while some of them show a weak break but no gravitationally dominated dense cores. Coupling those results with the physical characteristics of the clumps derived in this paper suggests a picture in which pressure confined clumps form prior to, or at higher rate compared to, the formation of gravitationally dominated dense cores in the clouds. This suggests that the formation of pressure confined clumps is a prerequisite for star formation, and introduces a natural threshold for star formation at $A_\mathrm{V}^\mathrm{tail}$.
     
      \item The star formation rate in the cloud complexes of our sample correlates strongly with the mass in the structures defined by the $A_\mathrm{V}^\mathrm{tail}$ threshold, as pointed out recently by \citet{lad10}, and furthermore, drops abruptly below that surface density \citep{hei10}. This supports the interpretation laid out in Item 3 above. Most importantly, the interpretation then provides a physically motivated explanation for the relation between star formation rate and the amount of dense material in the clouds reported by \citet{hei10} and \citet{lad10}.    
      \end{enumerate}

\begin{acknowledgements}
The authors would like to thank Mordecai-Mark Mac Low, Cornelis Dullemond, and Ralf Klessen for enlightening discussions regarding the topic. We would like to thank the anonymous referee for helping us to significantly improve the manuscript. We would also like to thank Jaime Pi\~neda for providing electronic material for Fig. \ref{fig:oph_co}. C.F.~has received funding from the European Research Council under the European Community's Seventh Framework Programme (FP7/2007-2013 Grant Agreement no.~247060) for the research presented in this work.
% NtS: COMPLETE only requires reference to Ridge et al.
\end{acknowledgements}

\end{document}